\documentclass[apj]{emulateapj}
\lefthead{TSUJIMOTO \& SHIGEYAMA }
\righthead{DIVERSITY OF  TYPE IA SUPERNOVAE}

\def\ltsima{$\; \buildrel < \over \sim\;$}
\def\ltsim{\lower.5ex\hbox{\ltsima}}
\def\gtsima{$\; \buildrel > \over\sim \;$}
\def\gtsim{\lower.5ex\hbox{\gtsima}}
\def\ms{$M_{\odot}$ }
\def\msp{$M_{\odot}$}

\slugcomment{Accepted for publication in ApJ Letters}

\begin{document}
\title{Diversity of Type Ia Supernovae Imprinted in Chemical Abundances}

\author{Takuji Tsujimoto\altaffilmark{1} and Toshikazu Shigeyama\altaffilmark{2}}

\affil{$^1$National Astronomical Observatory of Japan, Mitaka-shi,
Tokyo 181-8588, Japan; taku.tsujimoto@nao.ac.jp \\
$^2$Research Center for the Early Universe, Graduate School of Science, University of Tokyo, 7-3-1 Hongo, Bunkyo-ku, Tokyo 113-0033, Japan
}

\begin{abstract}
A time delay of Type Ia supernova (SN Ia) explosions hinders the imprint of their nucleosynthesis on stellar abundances. However, some occasional cases give birth to stars that avoid enrichment of their chemical compositions by massive stars and thereby exhibit a SN Ia-like elemental feature including a very low [Mg/Fe] ($\approx -1$). We highlight the elemental feature of Fe-group elements for two low-Mg/Fe objects detected in nearby galaxies, and propose the presence of a class of SNe Ia that yield the low abundance ratios of [Cr,Mn,Ni/Fe].  Our novel models of chemical evolution reveal that our proposed class of SNe Ia (slow SNe Ia) is associated with ones exploding on a long timescale after their stellar birth, and gives a significant impact on the chemical enrichment in the Large Magellanic Cloud (LMC). In the Galaxy, on the other hand, this effect is unseen due to the overwhelming enrichment  by the major class of SNe Ia that explode promptly (prompt SNe Ia) and eject a large amount of Fe-group elements. This nicely explains the different [Cr,Mn,Ni/Fe] features between the two galaxies as well as the puzzling feature seen in the LMC stars exhibiting very low Ca but normal Mg abundances. Furthermore, the corresponding channel of slow SN Ia is exemplified by performing detailed nucleosynthesis calculations in the scheme of SNe Ia resulting from a 0.8+0.6 \ms white dwarf merger. 
\end{abstract}


\section{Introduction}
Chemical abundance of long-lived stars is a powerful tool to probe nucleosynthesis in supernovae (SNe). In fact, detailed chemical compositions for very metal-poor stars in the Galaxy revealed by pioneering observations \citep[e.g.,][]{McWilliam_95, Cayrel_04, Yong_12} for more than a decade have brought a remarkable progress in the understanding of nucleosynthesis in core-collapse SNe, i.e., Type II SNe (SNe II) which first contribute to the chemical enrichment in the Universe. On the other hand, owing to a long time delay until the explosion, the nucleosynthesis products from Type Ia SNe (SNe Ia) are unavoidably superimposed on interstellar matter (ISM) already enriched by numerous SNe II, which prevents us from directly assessing SN Ia nucleosynthesis through an analysis of the observed data of stellar abundances. 

However, some circumstances will give birth to a pocket of SN Ia enriched gas, from which  stars inheriting their abundance pattern almost entirely from SNe Ia could be formed \citep{Tsujimoto_12a, Venn_12}. The most compelling relic of such an exclusive SN Ia enrichment must be a very low [Mg/Fe] (or [O/Fe]) since nucleosynthesis in SNe Ia predicts [Mg/Fe]$\approx -1.5$ \citep{Iwamoto_99} while individual SNe II normally give [Mg/Fe]$\approx$0.1-0.5 \citep{Cayrel_04}. \citet{Tsujimoto_12a} discuss the origin of the globular cluster (GC) NGC 1718 in the Large Magellanic Cloud (LMC) exhibiting [Mg/Fe]=$-0.9\pm0.3$ \citep{Colucci_12}, and found that its ratios of Fe-group elements (Cr, Mn, Ni) to Fe are identical to those predicted from the SN Ia nucleosynthesis. Besides NGC 1718, two stars are found in the Carina dwarf spheroidal (dSph) galaxy to have very low [Mg/Fe] ratios less than $-0.7$  \citep{Lemasle_12}. In addition to them, \citet{Venn_12} obtained  detailed elemental abundance of the star ({\it Car}-612) with [Mg/Fe]=$-0.5\pm$0.16 in the Carina, which is measured as [Mg/Fe]=$-0.9$ by \citet{Koch_08}. Surprisingly, {\it Car}-612 exhibits low [Fe-group/Fe] ratios such as [Cr/Fe], [Mn/Fe], [Ni/Fe] = $-0.20$, $-0.51$, $-0.46$  largely different from SN Ia-like ratios as in NGC 1718 showing [Cr/Fe], [Mn/Fe], [Ni/Fe] = $+0.25$, $-0.05$, $+0.09$.

In fact, these low [Fe-group/Fe] ratios are seen in stars for [Fe/H] \gtsim $-1$ in the LMC \citep{Pompeia_08}, which are a remarkable difference in the observed feature between the Galaxy and the LMC as shown in Figure 1. For instance, the [Cr/Fe] ratios among Galactic stars broadly reside within a narrow range of $-0.2$ to +0.1 while the LMC stars have [Cr/Fe] extending below $\sim -0.5$. We also see such low ratios in the Fornax dSph for Cr and Ni \citep{Letarte_10} and for Mn \citep{North_12}. These observations lead to two hypotheses that the elemental feature of {\it Car}-612 represents nucleosynthesis of a subclass of SNe Ia, and that these SNe Ia preferentially drive the chemical enrichment in dwarf galaxies.

This issue is emphasized by recent results regarding the delay time distribution (DTD) of SNe Ia yielded by the studies on the SN Ia rate in distant and nearby galaxies. These studies dramatically shorten the SN Ia's delay time, compared with its conventional timescale of $\sim$1 Gyr \citep{Pagel_95, Yoshii_96}. \citet{Mannucci_06} find that about 50 \% of SNe Ia explode soon after their stellar birth, and further works reveal that the DTD is proportional to $t^{-1}$ with its peak at around 0.1 Gyr extending to $\sim$ 10 Gyr \citep{Totani_08, Maoz_10}. 
Thus, the current view on SNe Ia is organized such that a majority of SNe Ia explode promptly after the bursting explosions of SNe II (prompt SNe Ia), and the rest gradually emerge with a long interval of Gyrs (slow SNe Ia).

To aim at unraveling two distinct chemical pathways of Fe-group elements seen in the Galaxy and the LMC,  we make two predictions: (i) nucleosynthesis of slow SNe Ia is characteristic of low [Fe-group/Fe] ratios as observed in {\it Car}-612, and (ii) the ejecta of prompt SNe Ia, which synthesize Fe-group elements more efficiently as already predicted in the existing SN models, can easily escape from the gravitational potential, if it is shallow,  owing to an inactive cooling in the low-density ISM after the bursting SN II explosions. Regarding prediction (i), it is crucial to validate whether some theoretical models for SNe Ia give low abundances for Fe-group elements, which will be discussed in \S 4. On the other hand, prediction (ii) implies that the total number of SNe Ia contributing to the chemical enrichment in dwarf galaxies should be smaller than that in the Galaxy. This is fully compatible with the implication from the observed high [Ba/Fe] ratios in the Fornax dSph and the LMC, which demands about one third of the SN Ia rate of the Galaxy \citep{Tsujimoto_11}.

In this Letter, we incorporate different nucleosynthesis yields from prompt and slow SNe Ia into the chemical evolution model, and present a unified scheme that explains the chemical feature among Fe-group elements for both the LMC and the Galaxy. It is further shown to give a nice explanation for the observed low-Ca feature associated with the relatively high Mg abundance in the LMC. We start with a brief review on the recent advancement in the understanding of theoretical models for SNe Ia (\S 2), followed by modeling the chemical evolutions for two galaxies (\S 3).  In \S 4, we perform one-dimensional numerical simulations for spherically symmetric white dwarfs (WDs) accreting matter composed of C+O to investigate detailed nucleosynthesis proceeding in merging WDs.

\section{Type Ia Supernova Progenitors}

From a theoretical point of view, SNe Ia might be composed of two types. One is the explosion of a WD with the mass of the Chandrasekhar limit accreting matter from the normal companion star \citep[single degenerate scenario:][]{1973ApJ...186.1007W,1984ApJS...54..335I} and the other is the explosion resulting from the merger of two WDs \citep[double degenerate scenario:][]{1984ApJ...277..355W,1984ApJS...54..335I}. The former scenario might be consistent with the similarity in light curves and spectra of SNe Ia in contrast to the variety of SNe II \citep{Hoeflich_93}, though there is no direct observational evidence to support this model \citep[e.g.,][]{2012Natur.481..164S}. The latter scenario naturally explains the observed diversity in SNe Ia by changing the combination of the masses of two WDs \citep{Pakmor_11}. A binary system of WDs with the secondary WD less massive than 0.6 \ms is associated with a  time scale longer than 1 Gyr. If the merger ignites C in the primary WD, this may explain dim SNe Ia in elliptical galaxies \citep{1989PASP..101..588F,2011AIPC.1379...17G} because the mass of the primary is likely to be comparable to that of the secondary. On the other hand, a pair of massive WDs both exceeding 1 \ms could explode as a SN Ia within 0.1 Gyr after their stellar  birth. 

Furthermore, \citet{2012Natur.481..164S} reported that they did not detect any stars to a visual magnitude limit of 26.9 in the central region of the supernova remnant SNR 0509-67.5 in the LMC, which is classified as Ia based on the spectrum of the light echo \citep{2008ApJ...680.1137R}. This observation seems to rule out single degenerate models and indicates that this particular supernova is the result of a violent merging of two WDs,  supporting the double degenerate scenario. A recent three-dimensional simulation done by \citet{2012ApJ...747L..10P} showed that a violent merging of two WDs  (1.1 \ms and 0.9 \msp) results in a supernova whose light curve and spectrum are in agreement with those of observed SNe Ia. Therefore numerical simulations have shown that the double degenerate model actually works. The question then arising is whether merging WDs synthesize heavy elements with the right abundance ratios observed in the star {\it Car}-612 as well as in the LMC.

\section{Chemical Evolution with prompt/slow SNe Ia}

\subsection{SN Ia models}

A novel ingredient in our chemical evolution model is the use of different nucleosynthesis yields for prompt and slow SN Ia. For the prompt SNe Ia, we adopt Fe-group-rich yields considered as a standard property of SN Ia nucleosynthesis. We use the delayed detonation model in a single-degenerate scenario from \citet{Iwamoto_99}, which predicts the [Fe-group/Fe] ratios identical to those of NGC 1718 as shown in Figure 1. The yields of this model are tabulated in Table 1 and incorporated into the chemical evolution model. For the slow SNe Ia, we assume a factor of $\sim$5 lower yields of Fe-group elements. We assign prompt SNe Ia to the first 1 Gyr in  the adopted DTD $\propto t_{\rm delay}^{-1}$ with the range of $0.1\leq t_{ \rm delay}\leq10$ Gyr so that they make up $\sim$70\% of SNe Ia. The DTD is normalized so that 8\% of the primary stars in binaries with the initial masses in the range of $3-8$\ms explode as SNe Ia \citep{Tsujimoto_12b}.

\subsection{Galaxy models}

\begin{figure}[t]
\vspace{0.2cm}
\begin{center}
\includegraphics[width=7cm,clip=true]{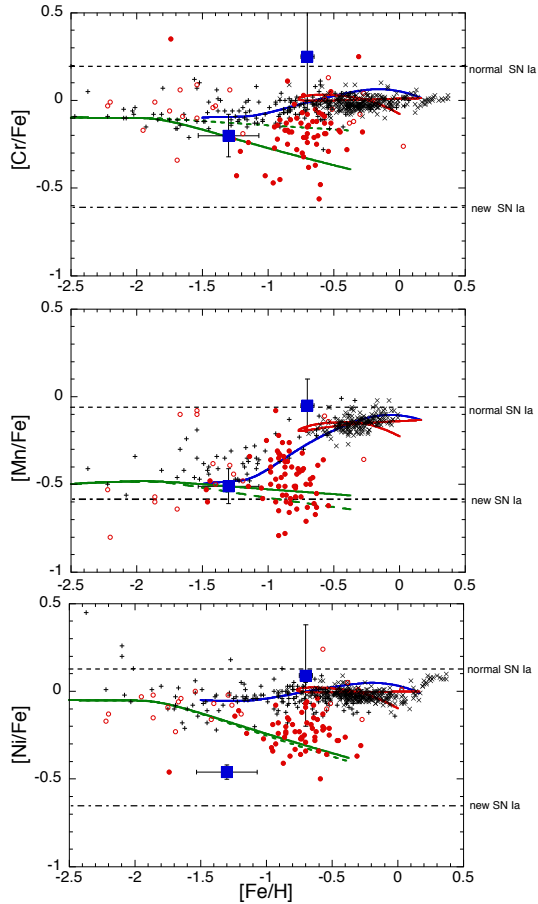}
\end{center}
\vspace{0.3cm}
\caption{Predicted [Cr,Mn,Ni/Fe] evolutions against [Fe/H] for the LMC (green lines, dashed ones for the cases with actual nucleosynthesis yields) and the Galaxy (blue lines for the thick disk and red lines for the thin disk), compared with the observations. For the observed data of the LMC, the GCs and field stars are denoted by open circles (Johnson et al.~2006; Mucciarelli et al.~2008, 2010; Colucci et al.~2012) and filled circles \citep[][but the Fornax dSph data by North et al. 2012 for Mn/Fe]{Pompeia_08}, respectively. For the Galaxy data, disk stars and halo stars are denoted by crosses \citep{Reddy_03, Bensby_05} and pluses \citep{Gratton_03}, respectively. The data of NGC 1718 and {\it Car}-612 are also shown by blue filled squares \citep{Colucci_12, Venn_12}.  Theoretical nucleosynthesis ratios in normal SNe Ia \citep{Iwamoto_99} and the newly proposed ratios are indicated by dashed lines and dash-dotted lines, respectively. 
}
\end{figure}

We construct models of chemical evolution for the Galaxy and the LMC. The Galaxy is modeled separately for the thick and thin disks, and each modeling is the same as in \citet{Tsujimoto_12b}. The LMC model is modified from the Galaxy's in respect of the star formation rate (SFR) and the initial mass function (IMF) \citep[see][]{Bekki_12}. The essence of these models is briefly described below.

{\sl the Galaxy}: First, the thick disk is formed rapidly, enriched by metal, and then the thin disk is gradually formed from the thick disk's debris gas mixed with the accreted metal-poor gas. We set the initial value of [Fe/H] to -1.5 which is implied from the metallicity distribution of thick disk stars. All SNe Ia contribute to chemical enrichment and a Salpeter IMF ($x$=$-1.35$) is adopted.

{\sl the LMC}: Star formation proceeds slowly, parameterized by a low SFR coefficient, $\nu$=0.15 Gyr$^{-1}$ (the mass fraction of the gas converted into stars per Gyr, and note that $\nu$=0.4, 2 Gyr$^{-1}$ for the thin and thick disk, respectively). It is assumed that only slow SNe Ia contribute to the chemical enrichment whereas the ejecta of prompt SNe Ia are all dispersed into the halo without being incorporated in the ISM. In addition, we reduce the contribution of heavy elements from SNe II by adopting a steep IMF ($x$=$-1.6$), which is predicted in particular from the Ba/Fe evolution \citep{Tsujimoto_11, Bekki_12}. This reduction of SN II enrichment mimics  a partial removal triggered by galactic winds associated with SNe II.

\subsection{Cr, Mn, Ni evolution}

Figure 1 shows the evolution of [Cr/Fe], [Mn/Fe], and [Ni/Fe] calculated for the thick disk (blue line), the thin disk (red line), and the LMC (green line). In the thick disk, the [Fe-group/Fe] ratios gradually increase owing to high Fe-group yields from prompt SNe Ia (dashed lines) for the first $\sim$1 Gyr and then take a slight downturn due to the contribution from slow SNe Ia (dashed-dotted line). Subsequently, the thin disk stars start to be formed from the thick disk's remaining gas with [Fe/H]$\sim$+0.2. First, [Fe/H] decreases owing to dilution by metal-poor infalling gas. This reverse evolution comes to an end when the chemical enrichment by star formation exceeds the effect of gas dilution and the usual evolution appears. On the other hand, in the LMC case, the [Fe-group/Fe] ratios decrease from a very low metallicity ([Fe/H]$\sim -1.8$) due to the absence of chemical enrichment by prompt SNe Ia. In reality, however, prompt SNe Ia are likely to contribute to the chemical enrichment at different degrees in different places in the LMC, which results in the observed large scatter broadly ranging between the lines calculated for the LMC and the Galaxy disks presented in Figure 1.

\subsection{Mg, Ca evolution}

\begin{figure}[t]
\vspace{0.2cm}
\begin{center}
\includegraphics[width=7cm,clip=true]{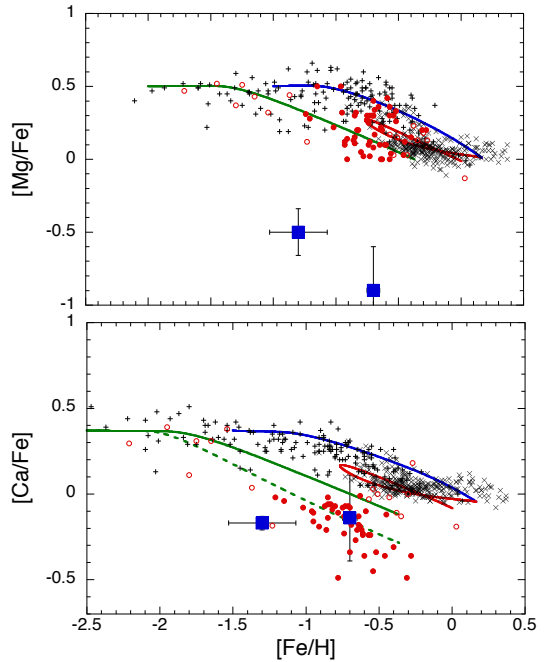}
\end{center}
\vspace{0.3cm}
\caption{Predicted and observed [Mg/Fe]-[Fe/H] (upper) and [Ca/Fe]-[Fe/H] (lower) correlations for both the LMC and the Galaxy. The symbols are the same as in Figure 1. The green dotted line is the prediction from the model in which the fraction of our proposed SNe Ia is set at twice the value adopted in standard model.
}
\end{figure}

For the LMC, we see a puzzling feature not only in the [Cr,Mn,Ni/Fe] trends but also in the differential trend between [Mg/Fe] and [Ca/Fe]. Though the common property for both is their low ratios in comparison with those of the Galaxy, the deficiency levels of Mg and Ca are quite different. [Mg/Fe] moderately decreases with most  values staying above the solar ratio, while [Ca/Fe] is severely low, reaching as low as [Ca/Fe]$\sim -0.5$. Such observed features can be consistently understood in the framework of our proposed SN Ia model. Since the Mg yield of SN Ia is completely negligible as compared with that of SN II, slow SNe Ia do not function to lower the Mg abundance. In contrast, the contribution of Ca from SNe Ia to chemical enrichment is large \citep[$\sim$30\% in the present-day local ISM;][]{Tsujimoto_95}. Therefore, if slow SNe Ia yield a low Ca abundance like Fe-group elements, [Ca/Fe] will become lower than [Mg/Fe] in the LMC. Here we assume that the Ca yield of slow SNe Ia is 1/5 that of prompt SNe Ia, keeping the same yield for Mg. The results for both the Galaxy and the LMC are shown in Figure 2. The predicted [Ca/Fe] in the LMC stars is not sufficiently lowered as a reflection of the low contribution of Ca from SNe Ia relative to SNe II as compared to the case of Cr, Mn, and Ni. Then, for instance, if we change the frequency ratio of prompt/slow from 0.7/0.3 to 0.4/0.6, the corresponding [Ca/Fe] evolution denoted by dashed line is in better agreement with the observation.

\section{Model for slow SNe Ia}

\begin{deluxetable*}{cccccc}
\tabletypesize{\scriptsize}
\tablecaption{Nucleosynthesis yields (\msp) of Fe-group elements in SNe Ia}
\tablehead{
\colhead{type} & \colhead{Fe} & \colhead{Cr} & \colhead{Mn} &
\colhead{Ni} & \colhead{comment}}
\startdata
\colhead{slow SN Ia} & 0.41 & $6.9\times10^{-3}$ & $1.1\times10^{-3}$ & $7.8\times10^{-3}$ & \colhead{nucleosynthesis calculation (this study)} \\
\colhead{} &  & $2.5\times10^{-3}$ & $1.5\times10^{-3}$ & $8.9\times10^{-3}$ & \colhead{prediction from chemical evolution$*$} \\
\hline
\colhead{prompt SN Ia} & 0.71 & $1.7\times10^{-2}$ & $7.1\times10^{-3}$ & $5.9\times10^{-2}$ & \colhead{WDD2 model in Iwamoto et al. (1999)} \\
\enddata
\tablecomments{*The Fe yield is assumed to be 0.63 \msp.}
\end{deluxetable*}

Here  we investigate if a WD merger can satisfy the nucleosynthesis conditions in slow SNe Ia required from the above chemical evolution model. 

\subsection{Model description}
As an initial model, an isothermal spherical WD is constructed by numerically integrating the equation of hydrostatic equilibrium from the center for a given central density. The used equation of state includes  electrons with arbitrary degeneracy, ideal ions, and thermal radiation. The temperature is assumed $10^8$ K to approximate the core temperature during the He-shell burning. The WD is embedded in a stationary accretion flow with a constant mass accretion rate. The velocity of the flow at the surface of WD is given by the free fall velocity. The flow is truncated where the total mass becomes the specified value.

We calculate the subsequent evolution of the flow with a one-dimensional Lagrangian PPM code including the  energy generation due to nuclear reactions and energy loss due to neutrino emissions \citep{Itoh96}. The code is a revised version used in \citet{1992ApJ...386L..13S}. The nuclear energy generation rate is evaluated by solving a nuclear reaction network including 13 $\alpha$-elements with the code developed by \citet{Mueller86}. The initial composition of matter is assumed to be half C and half O in this simplified nucleosynthesis calculation. The entire flow is assumed to be spherically symmetric. The mass coordinate is discretized with a uniform interval $\Delta M_r=2.5\times10^{-4}\,M_\odot$. The time interval $\Delta t$ is controlled by the Courant condition and the nuclear reaction timescale.

Using the thermal history of each grid point from the hydrodynamical calculation, we use a detailed nuclear reaction network involving 1266 nuclides up to $^{124}$Kr to obtain abundances of various elements \citep{Shigeyama10}. The initial abundances are assumed as follows. The mass fractions of elements heavier than O are 10\% of the solar. The remaining matter is half C and half O by mass. The reaction network equations are implicitly integrated.  The thermonuclear reaction rates in both nuclear reaction networks are taken from \citet{Angulo99} and \citet{Rauscher09}.

\subsection{Nucleosynthesis result}

\begin{figure}[t]
\vspace{0.2cm}
\begin{center}
\includegraphics[width=7.5cm,clip=true]{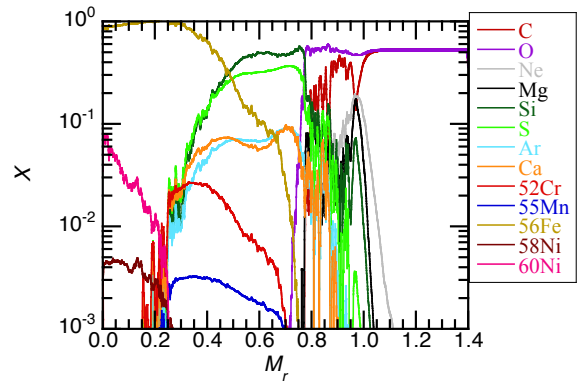}
\end{center}
\vspace{0.3cm}
\caption{Mass fractions of elements as a function of enclosed mass. This is a result of the explosion of a 0.8 \ms C+O WD accreting 0.6 \ms composed of C+O at the rate of 0.07 \ms yr$^{-1}$.
}
\end{figure}

The intense accretion generates two shock waves propagating outward in the accreted matter and inward in the WD. The shocked matter incinerates C and O up to Si and S near the contact surface. The subsequent rarefaction waves prohibits further nuclear reactions. When the shock generated in the star eventually arrives at the center without igniting C or O along the way, it ignites these elements at the center and generates a detonation shock wave.  This shock propagates outward and incinerates a significant fraction of C and O in the star up to Fe-group elements. 

Here, we show the results of the explosion of a 0.8 \ms WD accreting 0.6 \ms of a mass accretion rate of 0.07 \ms s$^{-1}$ (Fig.~3). This yields 0.41 \ms Fe as shown in Table 1, which is a factor 1.5 smaller than the Fe masses from  prompt SNe Ia. A 0.6 \ms WD is the final product of a $\sim$2 \ms star \citep{2006MNRAS.369..383D, Salaris_09}, and its lifetime is about 1 Gyr. Therefore, taking into account the time it takes for the orbital decay of merging WDs due to the gravitational radiation, this explosion results in a dim SN Ia with a delay time $>$ 1 Gyr. 

The resulting abundance ratios of Mn/Fe and Ni/Fe in the ejecta meet the requirement from the chemical evolution model. On the other hand, the model shows overabundance of Cr as compared to the predicted value, but Cr/Fe is reduced by $\sim$30 \% from that of the prompt SN Ia and becomes closer to the ratio predicted from the chemical evolution model. The results of chemical evolution models adopting these actual nucleosynthesis yields are shown for the LMC case by dashed green lines in Figure~1. The amount of Ca from this model is 0.032 \msp, which results in a Ca/Fe ratio comparable to that from prompt SNe Ia. All these features are inherent in the spherical detonation of a WD with the mass significantly lower than the Chandrasekhar limit.

In fact, Ca and Cr are synthesized in the outer layers of the primary WD. Since the actual accretion flow forms a disk, the assumption of spherical accretion overestimates the compression in the primary star. The resulting  shallower densities might inhibit synthesis of elements heavier than S, which is realized in the accreted flow of the present spherical model at $0.8 M_\odot < M_r <1.0 M_\odot$ in terms of the enclosed mass coordinate $M_r$ (see Fig.~3). Quantitative arguments need to wait for detailed nucleosynthesis calculations based on three-dimensional simulations like \citet{Pakmor_10,Pakmor_11}.

\section{Conclusions}

We have assessed the diversity of SNe Ia yielding different nucleosynthesis products from two theoretical approaches; (i) modeling the chemical evolution of two galaxies exhibiting distinct chemical features, i.e., the Galaxy and the LMC, and (ii) performing  nucleosynthesis calculations in the scheme of the double degenerate scenario. Accordingly, we propose a subclass of SNe Ia which releases  low abundances of Fe-group elements with a long timescale and which leaves little relics in the chemical abundances of the Galaxy. The presence of dim SNe Ia has been already predicted as a result of the merger of two WDs \citep{Pakmor_10,Pakmor_11}. In dwarf galaxies, however, it is difficult to trap the ejecta from prompt SNe Ia that synthesize a large amount of Fe-group elements and explode in a burst. Hence, the proposed slow SNe Ia characterize the chemical features of a late evolution which leads to low ratios of [Cr,Mn,Ni/Fe]  and to a large scatter in these ratios. Our scenario should be validated by further studies both observationally and theoretically, in particular by  determination of detailed elemental abundances for more stars exhibiting very low [Mg/Fe] together with three-dimensional simulations for SN Ia nucleosynthesis.

\acknowledgements
The authors wish to thank an anonymous referee for his/her valuable comments that has considerably improved the paper.

\end{document}